\documentclass[aps,prb,superscriptaddress,showpacs,reprint]{revtex4-1}

\usepackage[utf8]{inputenc}
\usepackage{amsmath,amsfonts,amssymb}
\usepackage{pifont}
\usepackage{rotating}
\usepackage{enumerate}
\usepackage{ulem}
\usepackage{graphicx}    
\usepackage{epstopdf}
\usepackage{color}
\usepackage{soul} 
\usepackage{subfigure}
\usepackage{multirow}
\usepackage{bm}

\usepackage{hyperref}
\definecolor{dark-red}{rgb}{0.9,0.15,0.15}
\definecolor{dark-blue}{rgb}{0.15,0.15,0.4}
\definecolor{medium-blue}{rgb}{0,0,0.5}
\hypersetup{
    colorlinks={true},
     linkcolor={medium-blue},
    citecolor={blue},
     urlcolor={medium-blue}
}

\begin{document}
\title{CoFeVSb: A Promising Candidate for Spin Valve and Thermoelectric Applications}
\author{Jadupati Nag}
\affiliation{Department of Physics, Indian Institute of Technology Bombay, Mumbai 400076, India}

\author{Deepika Rani}
\affiliation{Department of Physics, Indian Institute of Technology Bombay, Mumbai 400076, India}
\affiliation{Department of Physics, Indian Institute of Technology Delhi, New Delhi, 110016, India}

\author{Durgesh Singh}
\affiliation{Department of Physics, Indian Institute of Technology Bombay, Mumbai 400076, India}

\author{R. Venkatesh}
\affiliation{UGC-DAE Consortium for Scientific Research, University Campus, Khandwa Road, Indore-452001, India}

\author{Bhawna Sahni}
\affiliation{Department of Physics, Indian Institute of Technology Bombay, Mumbai 400076, India}

\author{A. K. Yadav}
\affiliation{Atomic $\&$ Molecular Physics Division, Bhabha Atomic Research Centre, Mumbai – 400094, India}

\author{S. N. Jha}
\affiliation{Atomic $\&$ Molecular Physics Division, Bhabha Atomic Research Centre, Mumbai – 400094, India}

\author{D. Bhattacharyya }
\affiliation{Atomic $\&$ Molecular Physics Division, Bhabha Atomic Research Centre, Mumbai – 400094, India}

\author{P. D. Babu}
\affiliation{UGC-DAE Consortium for Scientific Research, Mumbai Centre, BARC Campus, Mumbai 400085, India}

\author{K. G. Suresh}
\email{suresh@phy.iitb.ac.in}
\affiliation{Department of Physics, Indian Institute of Technology Bombay, Mumbai 400076, India}

\author{Aftab Alam}
\email{aftab@phy.iitb.ac.in}
\affiliation{Department of Physics, Indian Institute of Technology Bombay, Mumbai 400076, India}


\begin{abstract}
We report a combined theoretical and experimental study of a novel quaternary Heusler system CoFeVSb from the view point of room temperature spintronics and thermoelectric applications. It crystallizes in cubic structure with small DO$_3$-type disorder. The presence of disorder is confirmed by room temperature synchrotron X-ray diffraction(XRD) and extended X-ray absorption fine structure (EXAFS) measurements. Magnetization data reveal high ordering temperature with a saturation magnetization of 2.2 $\mu_B$/f.u. Resistivity measurements reflect half-metallic nature. Double hysteresis loop along with asymmetry in the magnetoresistance(MR) data reveals room temperature spin-valve feature, which remains stable even at 300 K. Hall measurements show anomalous behavior with significant contribution from intrinsic Berry phase. This compound also large room temperature power factor ($\sim0.62$ mWatt/m/K$^{2}$) and ultra low lattice thermal conductivity ($\sim0.4$ W/m/K), making it a promising candidate for thermoelectric application. Ab-initio calculations suggest weak half-metallic behavior and reduced magnetization (in agreement with experiment) in presence of DO$_3$ disorder. We have also found an energetically competing \textcolor{black}{ferromagnetic (FM)/antiferromagnetic (AFM)} interface structure within an otherwise FM matrix: one of the prerequisites for spin valve behavior. Coexistence of so many promising features in a single system is rare, and hence CoFeVSb gives a fertile platform to explore numerous applications in future.
\end{abstract}

\date{\today}

\maketitle

{\it{Introduction:}}
Heusler alloys have drawn enormous attention in several fields due to their interesting properties. Most of them are versatile and quite promising because of their robust structure, high Curie temperature {($T_C$)}, and high spin polarization; hence highly suited for spin-transport applications.
Half metallic ferromagnets (HMFs),\cite{PhysRevLett.50.2024,pickett2001half} which are one of the earliest classes of spin-polarized materials, have led a revolution
in the field of spintronics. HMFs possess a unique electronic structure exhibiting a finite band gap in one spin channel and metallic behavior in the other.
There has been a large number of materials in the Heusler family,\cite{bombor2013half,GRAF20111,PhysRevB.96.184404,moodera1994nature} which are reported to be HMF 
and are predicted to show 100\% spin-polarization because of the complete absence of minority spin states at the Fermi level ($E_F$).
Such materials are in great demand for applications such as magnetic tunnel junctions\cite{tezuka2006tunnel,sakuraba2010co} spin injectors, 
spin valves \cite{kasahara2014greatly,furubayashi2013temperature} etc.  A spin valve is a device, consisting of two or more conducting magnetic layers, where electrical resistance 
can change between two values depending on the relative alignment of the magnetization in the layers. The resistive state of such a composite system can be switched by 
changing the direction of the applied magnetic field, resulting in asymmetric magnetoresistance (MR). The spin valve concept is
often used in devices such as magnetic sensing, recording etc. This feature is generally observed in
multilayer ferromagnetic thin films \cite{PhysRevB.43.1297} and is extremely rare in bulk materials.\cite{PhysRevLett.109.246601}  
Another important research area where Heusler alloys, specially Half Heuslers, are found promising is the thermoelectric(TE) application.\cite{sakurada2005effect,yan2011enhanced}
 
In this article, we report a combined theoretical and experimental study on promising spin-valve and TE properties of a quaternary Heusler alloy, CoFeVSb (CFVS). Though there 
exists a theoretical study on ordered CFVS\cite{xiong2014search} predicting its half metallic behavior, no experimental/theoretical investigation reporting its potential for spin valve and TE applications is
reported so far. Using combined synchrotron XRD and EXAFS measurements, we confirm CFVS to crystallize in
LiMgPdSn structure with robust partial DO$_3$ disorder. Our ab-initio calculations confirm half-metallic nature of CFVS in its ordered phase, as reported earlier. The disordered DO$_3$ structure, however, makes it a weak half metal, as also supported by our magnetotransport measurement. The measured saturation magnetization is found to be 2.2 $\mu_B$/f.u. along with a high $T_C$, in fair agreement with theory. Magnetization and MR data strongly indicate room temperature (RT) spin valve feature in this alloy. A careful analysis of Hall data indicates
the dominance of intrinsic Berry phase contribution to the anomalous Hall effect (AHE). CFVS also shows enormous promise for TE applications, with a high thermopower ($\sim0.62$ mWatt/m/K$^{2}$) and ultra low simulated lattice thermal conductivity ($\sim0.4$ W/m/K) at RT.
High $T_C$, robust structure, half metallicity, promising spin-valve and thermoelectric properties make CFVS a potential candidate for multifunctional applications.

{\it{Experimental and Computational Details:}}
Experimental details including synthesis procedure, normal and synchrotron XRD, EXAFS, magnetization and transport measurements are given in supplementary material (SM).\cite{supplement} Computational details are also provided in SM.\cite{supplement}

\begin{figure}[t]
\centering
\includegraphics[width=1.0\linewidth]{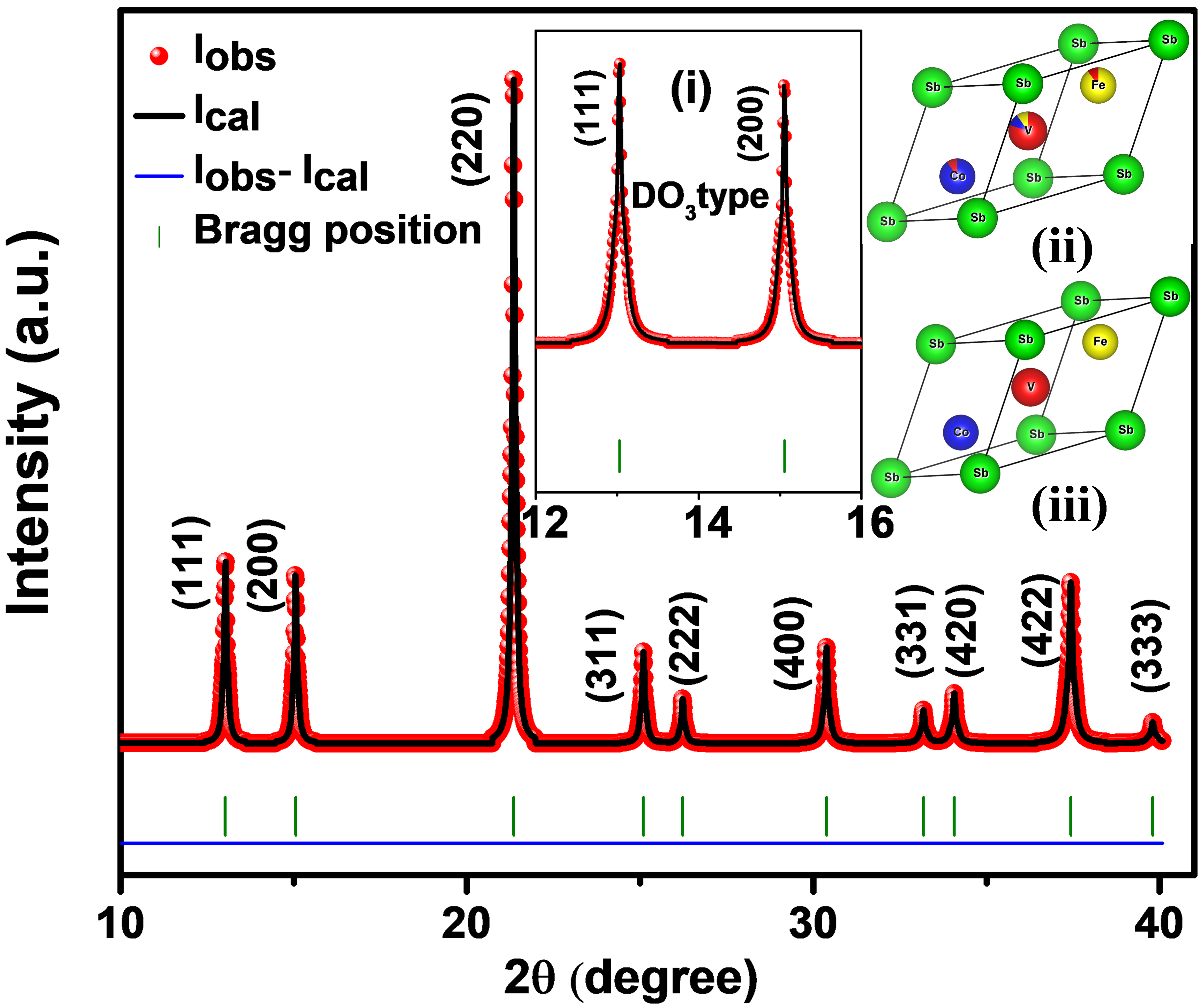}
	\caption{For CoFeVSb, synchrotron XRD pattern at RT including the Rietveld refined data for type I configuration with partial DO$_3$ disorder. Inset (i) shows a zoomed-in view near (111) and (200) peaks. Insets (ii,iii) display the crystal structures for partial DO$_3$ disorder and pure Y-type.}
\label{fig:xrd-CFVS}
\end{figure}

{\it Crystal Structure (X-ray diffraction):}
CFVS crystallizes in the LiMgPdSn prototype structure (space group $F\bar{4}3m$) with lattice parameter 5.82 \AA.
There exist three non-degenerate crystal configurations for any equi-atomic quaternary Heusler structure.\cite{PhysRevB.104.134406} These are discussed in theoretical results section, which confirm Type-I to be the energetically most stable configuration.  Figure \ref{fig:xrd-CFVS} shows Rietveld refined ($\chi^2$ = 0.5) synchrotron XRD (SXRD) pattern (using $\lambda = 0.6525\ \AA $) at RT, where Type-I configuration with 12.5\% disorder between Co-V (X-Y) and 12.5\% disorder between Fe-V (X$'$ -Y) atoms is considered for fitting. This combination gives the best fitting compared to all other ordered and disordered ones we tried. Normal XRD data show a similar trend for Rietveld refinement, see SM.\cite{supplement}
Inset (i) of Fig. \ref{fig:xrd-CFVS} shows a zoomed in view of the the refined SXRD data near (111) and (200) peaks.
As these superlattice reflections are very prominent, possibility of  A2 or B2-type disorder
is ruled out. There may be some possibility of DO$_3$ disorder because of a slight reduction of (200) peak intensity in comparison to (111) peak. This is also predicted by our EXAFS data, described below. In general, for XX$^{'}$YZ alloy, intermixing of X/X$'$ with Y or X/X$'$ with Z sites yields DO$_3$ type disorder. In the present case, we have considered only the swap disorder involving X/X$'$ and Y (Co/Fe with V) since EXAFS measurements have not found any contribution from swapping of Sb with Co/Fe.

{\it Crystal Structure (EXAFS):}
Figure \ref{fig:exafs-CFVS} shows the XANES spectra of CFVS at Co, Fe and V K-edges along with those of the respective metal foils. These K-edges coincide with the edges of their respective metal foils, indicating the absence of oxide phases. The oscillations in the spectra immediately above the absorption edge are completely different as compared to their respective foils for Co and V K-edges. 
The $k^2$ weighted $\chi (k)$ vs. $k$ spectra are shown in the insets of Fig. \ref{fig:exafs-CFVS}(d,e,f).
\begin{figure}[t]
\centering
\includegraphics[width= 8.8cm,height=7.5cm]{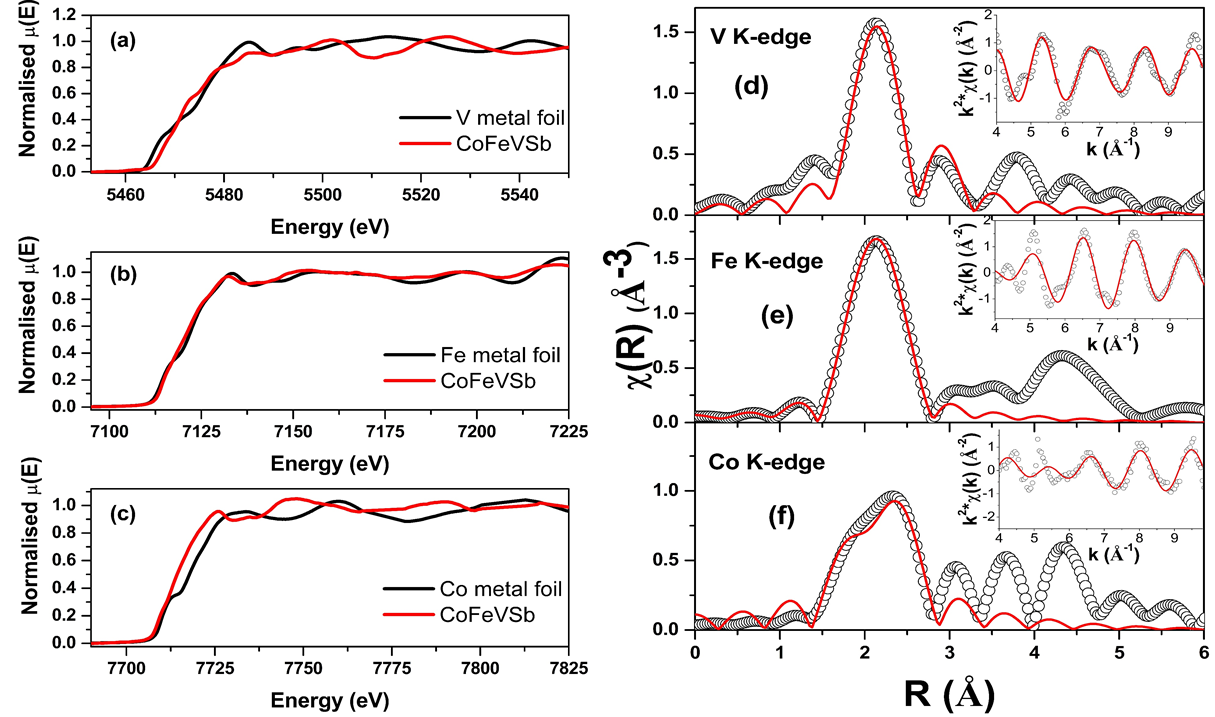}
	\caption{For CoFeVSb, normalised XANES spectra at (a)V K-edge, (b)Fe K-edge and (c)Co K-edge along with that of the corresponding metal foil. (d-f) Fourier transformed EXAFS spectra at (top to bottom)  V, Fe and Co K-edges (scatter points) along with the theoretical fit (solid line). The respective insets show raw and fitted data in wave vector($k$) space.}
\label{fig:exafs-CFVS}
\end{figure}
The $\chi(R)$ vs. $R$ spectra (generated from the Fourier transform of $k^2 \chi(k)$) are shown in Fig. \ref{fig:exafs-CFVS}(d,e,f) measured at Co, Fe and V K-edges. Figure \ref{fig:exafs-CFVS}(d-f) shows the best fit $\chi(R)$ vs. $R$ spectra, along with the experimental data. The anti-site disorder paths have been generated by swapping the absorption atom site. The bond distances, disorder (Debye-Waller) factor ($\sigma^2$ ) which gives the mean square fluctuations in the distances, and the fraction of anti-site disorder ($x$) are used as the fitting parameters (given in Table S1 of SM\cite{supplement}). The first coordination peak is used for fitting because of the relatively low data range $k=3-10\ \AA^{-1}$. From the fitting, it appears that there is no contribution of anti-site disorder due to the swapping of Sb with Co and Fe, rather it arises due to the swapping of V with Fe/Co. Thus, EXAFS measurements confirm the possibility of anti-site disorder between Co-V and Fe-V sites, giving rise to the DO$_3$ disorder, which is also supported by XRD data.
\begin{figure*}[t]
\centering
\includegraphics[width=\linewidth]{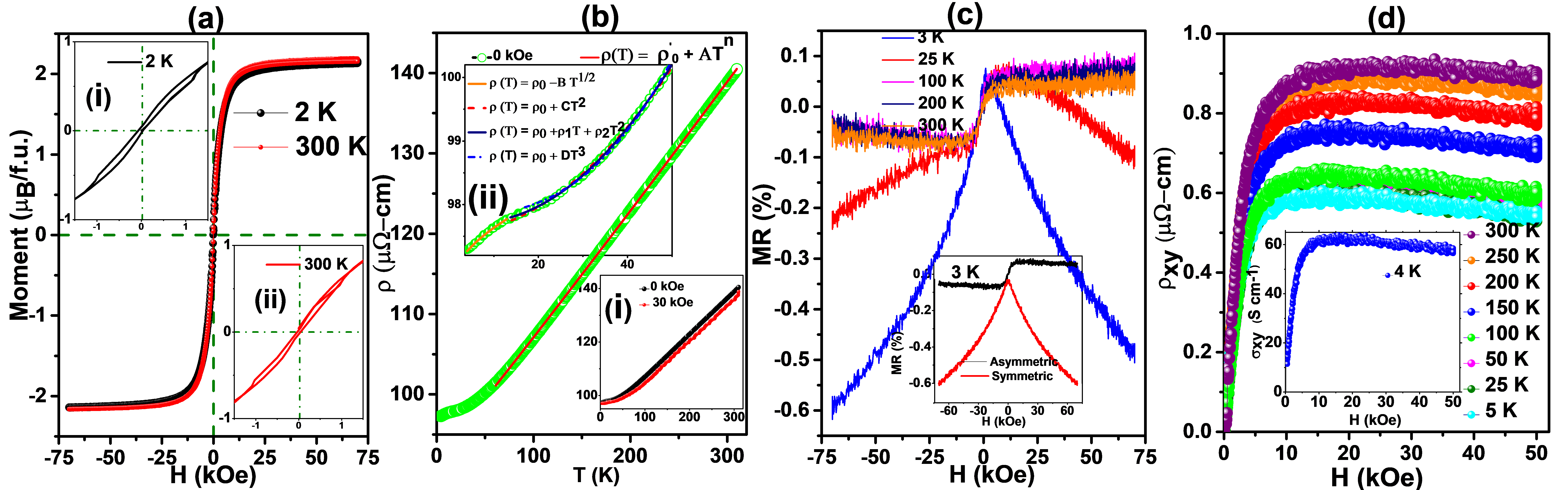}
\caption{For CoFeVSb, (a) M vs. H curves at 2 and 300 K. The insets display zoomed-in view (b) Longitudinal resistivity($\rho$) vs. T, along with the power law fitting in high T range. Inset (i) shows $\rho$ vs. T at 0 and 30 kOe fields while (ii) shows zoomed in view of the low T behavior, along with various fitting models. (c) MR vs. H at various T. Inset shows the asymmetric and symmetric components at 3 K. (d) Hall resistivity($\rho_{xy}$) vs. H at different T. Inset shows the Hall conductivity vs. H at 4 K.}
\label{fig3}
\end{figure*}

{\it Magnetic Properties:}
Figure \ref{fig3}(a) shows M vs. H curves for CFVS. Insets (i and ii) show a zoomed in view at 2 K and 300 K. The M vs. T curve is shown in Fig. S5 of SM\cite{supplement} which indicates high $T_C$ for this alloy. M-H curves show plateau-like magnetization with a small but non-zero hysteresis and 70 Oe coercivity (at 3 K), indicating CFVS to be a soft ferromagnet. The low field part of the M-H curve shown in the insets clearly indicate the presence of a double hysteresis loop, a property usually signifying the spin valve effect.\cite{yoneda2015simulation} The plateau-like feature in the M-H curves seen here is nearly identical to that seen in some superlattice structures involving magnetic oxides. Such a behavior is attributed to the coexistence of spin-glass and the ferromagnetic background in compounds with magnetic anisotropy.\cite{pang2017spin}. One can also notice a step-like feature at about 1 kOe, indicating the presence of an additional, low magnetization phase. \textcolor{black}{One should also note that total magnetic moment at 300 K is slightly higher than 2K, that again indicates the possibility of small ferri/AFM regimes.}
The total moment ($m$) per formula unit for a Heusler alloy can be calculated in terms of the number of valence electrons ($N_v$),\citep{graf2011simple} using Slater-Pauling (SP) rule:
$m = (N_v - 24) \ \ \ \mu_B/f.u.$\cite{ozdougan2013slater,zheng2012band}
As such, for CFVS, a moment of 3.0 $\mu_B$/f.u. is expected in its completely ordered state. The measured moment turns out to be 2.2 $\mu_B$/f.u. This difference is attributed partially to the anti-site (DO$_3$) disorder \cite{vidal2011influence} \textcolor{black}{or the presence of competing magnetic ordering.  
This is also supported by significant contrast in our magnetic force microscopy (MFM) data highlighting boundaries between domains of two competing magnetic phases even at RT (see Fig. S2 and experimental details in SM\cite{supplement}). } All these coexisting features, such as double hysteresis loop, reduced magnetization, metamagnetic-like step in the M-H loop, \textcolor{black}{contrasting regions of competing magnetic phases in MFM data strongly indicates the existence of some disordered phase involving FM/AFM clusters within a FM matrix.} Such features are observed rarely in a bulk material and are responsible for the spin-valve effect.

{\it Transport properties (Resistivity):}
Figure \ref{fig3}(b) shows the temperature dependence of longitudinal resistivity ($\rho_{xx}$) at different fields (0 and 30 kOe) (see inset (i)). $\rho_{xx}$ increases and varies almost linearly with T towards 300 K. 
To get a better understanding of the variation of resistivity with T, we analyzed the data in three different T ranges:
I (3 $<$T$<$13 K), II (14 $<$T$<$60 K) and III (60$<$T$<$310 K). \textcolor{black}{The details of the fitting considering different models are given in SM.\cite{supplement}} From the fitting, it appears that T$^{2}$ dependence is very weak and linear behavior dominates throughout the T-range (specially high T). Such a dependence implies the dominance of electron-phonon scattering. This, in turn, hints towards the redundant spin-flip scattering, arising out of single-magnon scattering, suggesting the HMF behavior of CFVS.\cite{bombor2013half}

{\it Transport properties (Magnetoresistance):}
Figure \ref{fig3}(c) shows the field dependence of MR, defined as MR(H)=$ \left[ \rho(H) - \rho(0)\right]/\rho(0)$, at various T. Magnitude of MR is low, increases with H in a sub-linear fashion and does not saturate till 70 kOe at 3 K. Interestingly, it shows an asymmetric feature between positive and negative fields. Inset of Fig. \ref{fig3}(c) shows the asymmetric and symmetric components of MR at 3 K.\footnote{The asymmetric component is calculated as MR(H)=$ \left[ M(H) - M(-H)\right]/2$ while the symmetric component as MR(H)=$ \left[ M(H) + M(-H)\right]/2$.}
The symmetric part is -ve and shows a linear variation with H. As T is raised to 25 K, in addition to the asymmetry, one can notice a clear hump as H changes from negative to positive. At 25 K, a sharp plateau like feature is observed at lower H, which gets suppressed at higher H giving rise to the asymmetric nature. MR exhibits a crossover behavior (-ve to +ve) starting from 50 K, which continues till 300 K. These features of MR clearly reflect the spin-valve like behavior,\cite{PhysRevLett.109.246601} which remains intact in the T-range of 50-300 K and even under a field of 70 kOe.\cite{agarwal2018spin} One should note that the double hysteresis loop also points towards the spin valve like feature in this alloy. The emergence of competing magnetic phases necessary for the spin-valve effect is discussed in the theoretical results section.

{\it Transport properties (Hall Measurements):}
Figure \ref{fig3}(d) shows the Hall resistivity $\rho_{xy}$ vs. H, at different T. The H-dependance of Hall conductivity ($\sigma_{xy}$ $\approx$ $\frac{\rho_{xy}}{\rho_{xx}^2}$) is shown in the inset. The residual $\sigma_{xy0}$ is found to be 64.4 S-cm$^{-1}$ at 4 K, which is much smaller compared to that of half-metallic systems.\cite{bombor2013half} $\rho_{xy}$ mainly consists of two contributions, (i) ordinary Hall effect (OHE) and (ii) anomalous Hall effect. The later arises due to the magnetic contribution as a result of the asymmetric scattering of conducting electrons.


In general, Hall resistivity can be expressed as,\cite{RevModPhys.82.1539}
$\rho_{xy}(T)=\rho_{xy}^{OHE} + \rho_{xy}^{AHE}=R_{0}H+R_{A}M$,
where $\rho_{xy}^{AHE}$, $R_0$, $R_A$ and $M$ are anomalous Hall resistivity, ordinary Hall coefficient, anomalous Hall coefficient and magnetization respectively. In low H-range, AHE dominates while OHE is dominant at relatively higher H. AHE contributions to $\rho_{xy}$ can be obtained by extrapolating the high field data to zero field. Further details of AHE analysis are given in SM.\cite{supplement}  
The negative slope of ordinary Hall coefficient suggests electrons to be majority charge carriers, an inference that matches with the TE results, shown below. From the fitting of Hall data and detailed AHE analysis (see SM\cite{supplement}) we obtain a non-zero Karplus-Luttinger term  confirming a finite contribution of intrinsic mechanism. This, in turn, indicates the dominance of
intrinsic Berry phase contribution to AHE.\cite{PhysRevLett.103.087206}

\begin{figure}[t]
\centering
\includegraphics[width=8.5cm,height=5cm]{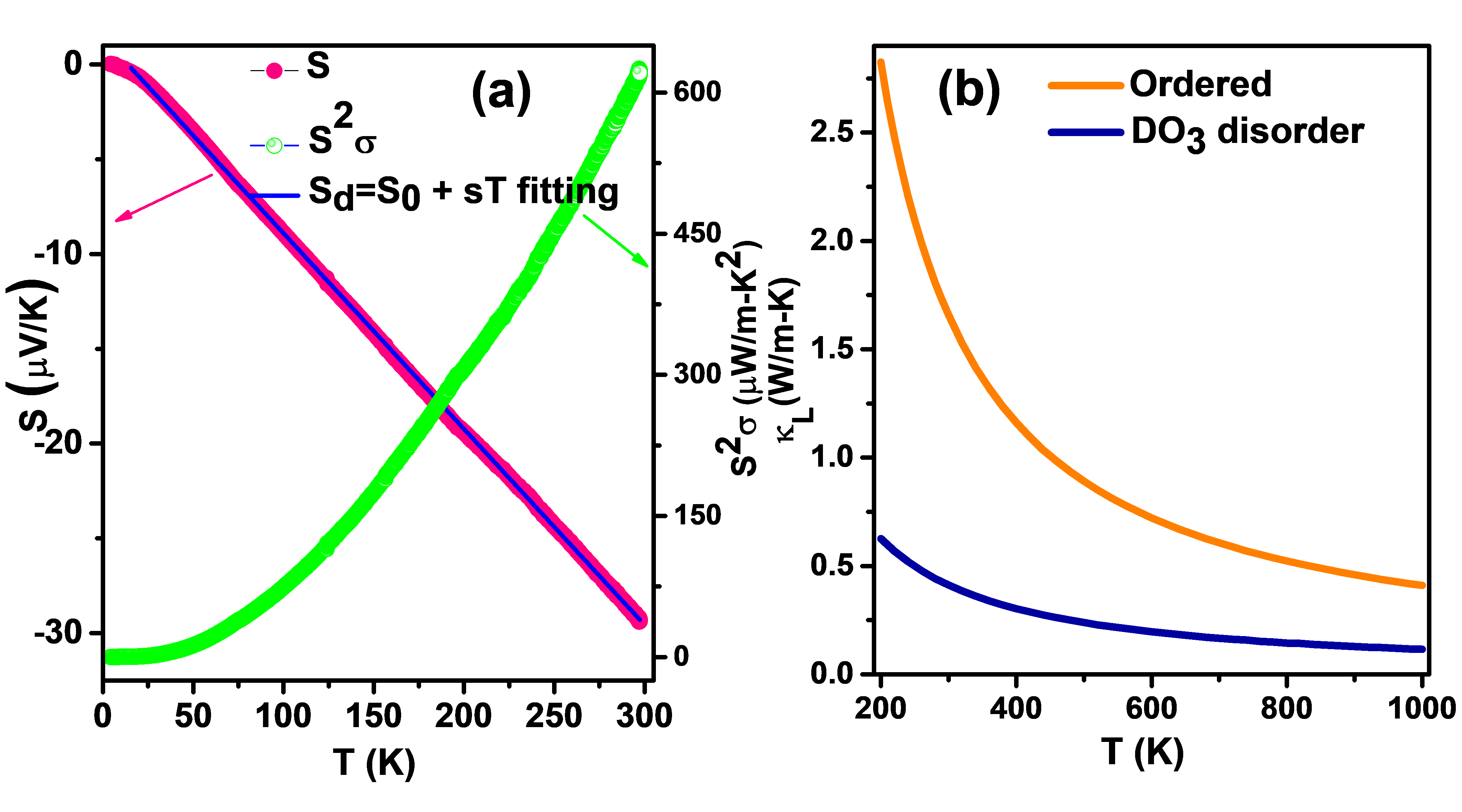}
\caption{For CoFeVSb, (a) Seebeck coefficient ($S$) and power factor ($S^2\sigma$) vs. T. The fitting to the $S$ vs. T curve illustrates the linear behavior of diffusion thermopower (S$_d$=S$_0$+s$T$) between 20-300 K. (b) Simulated lattice thermal conductivity($\kappa_L$) for ordered and DO$_3$ disordered structure.}
\label{fig:CFVS-TEP}
\end{figure}

{\it Thermoelectric properties: }
Figure \ref{fig:CFVS-TEP}(a) shows the variations of the Seebeck coefficient (S) and power factor ($S^2\sigma$) with T. S increases linearly with T throughout the T-range, which is a typical behavior of weak half-metal.
The negative slope of S vs. T plot corresponds to a purely electron driven $S$, which is reflected in the Hall data as well. The linear behavior of $S$ suggests the dominance of diffusion thermopower. The magnitude of
S is 30 $\mu V/K$ at 300 K, which is fairly high and comparable to other potential TE materials, at RT.\cite{yu2009high,hayashi2017structural,lue2002thermoelectric}
Power factor shows an almost quadratic nature with a maximum value of 621 $\mu Watt/ m-K^2$ at RT. This is one of the highest among many Heusler based TE materials,\cite{sakurada2005effect,yan2011enhanced} and also in line with other reported promising TE materials.\cite{fu2015realizing,huang2015new,fu2013electron}
Carrier concentration ($n$) is evaluated by fitting the $S$-data with the equation S$_d$=S$_0$+s$T$ in
the T-range of 25-300 K, where S$_d$ is the diffusion thermopower and $S_0$ is a constant. s = $\frac{\pi^2{k_B}^2}{eE_{F}}$ yields a carrier concentration of $n \simeq {10^{21}}$. Typical $n$ for promising TE materials lie in the range 10$^{19}$ to 10$^{21}$ cm$^{-3}$.


To further evaluate the potential of CFVS for TE applications, we have simulated the lattice thermal conductivity ($\kappa_L$) with and without DO$_3$ disorder, as shown in Fig. \ref{fig:CFVS-TEP}(b). CFVS with DO$_3$ disorder (the actual experimental phase) shows ultra low $\kappa_L$ values (0.1-0.6 Wm$^{-1}$K$^{-1}$) due to the increased phonon scattering in comparison to the completely ordered phase. Similar $\kappa_L$ values are reported in a few other promising TE materials as well.\cite{sajjad2020ultralow,mukhopadhyay2018two} This again confirms the potential of CFVS for TE applications.

{\it Theoretical Results: }
Ab-initio total energy calculations are performed considering three different structural configurations (Type-I, Type-II and Type-III) of CFVS (see SM\cite{supplement} for more details) We also simulated various magnetic
states by considering different magnetic arrangements (antiferro-, ferro- and ferri-magnetic) in all the three configurations.
Table SIII of SM\cite{supplement} shows the energetics, optimized lattice parameters and moments of the three ordered configurations. Type-I turns out to be energetically the most stable configuration with ferromagnetic ordering.
Figure \ref{fig:CFVS-band} displays the spin polarized band structure and density of states (DoS) for this configuration, which clearly shows a half-metallic nature with a band gap of 0.48 eV in the minority spin channel.


\begin{figure}[t]
\centering
\includegraphics[width= 0.98\linewidth]{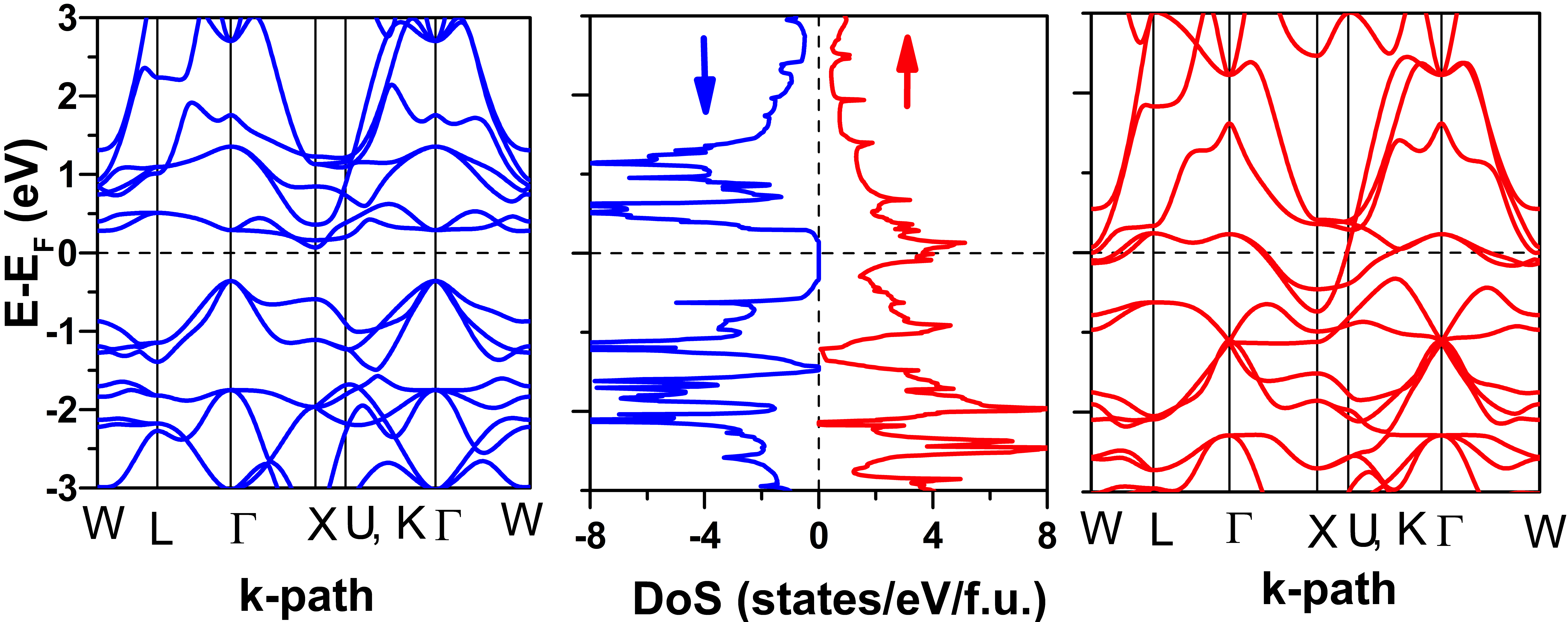}
\caption{ Spin polarized band structure and density of states of Type I configuration of pure CoFeVSb at $a_0$=5.9 \AA.}
\label{fig:CFVS-band}
\end{figure}
\par
The total magnetic moment of Type-I configuration is 3.0 $\mu_B$/f.u., which agrees well with the SP rule. In contrast, the experimentally observed moment is 2.2 $\mu_B$/f.u. This difference is partially due to the DO$_3$ disorder in this system, as observed experimentally.
To understand this discrepancy, we have constructed a DO$_3$ disordered structure of CFVS by swapping 1 Co with 1 V and 1 Fe with 1 V atoms in a $2\times2\times2$ supercell, which gives a 12.5\% swap disorder, as predicted by XRD and EXAFS measurements. See \footnote{XRD and EXAFS measurements reveal a DO$_3$-type disorder in CFVS, which in the present case arises due to disorder between Co and V as well as Fe and V sites. To simulate the experimentally observed structure, a $2\times2\times2$ supercell of the primitive cell of the most stable ordered configuration (Type I) of CFVS was constructed. This supercell contains a total of 32 atoms, including 8-atoms of each kind. In a $2\times2\times2$ supercell, exchanging one of the eight Co/Fe atom positions with one of the eight 'V' atom positions leads to a 12.5\% swap disorder between Co/Fe and V atoms. All possible configurations for replacement of Co/Fe by V and vice versa were simulated, and the results of energetically most stable configurations are chosen to present here.  As compared to the completely ordered structure, the atom projected local moments in the disordered case changes considerably due to the change in the local atomic environment.} for more details about this structure. The simulation of this structure indeed gives a reduced magnetization of 2.5 $\mu_B$/f.u., in line with experimental prediction. With DO$_3$ disorder, the V atoms which replace Co and Fe sites remain FM but gain moment as compared to their own designated Wyckoff sites. In contrast, Co and Fe atoms which replace V sites become antiferromagnetically aligned and hence cause an overall reduction in the net moment. Also, with DO$_3$ disorder, CFVS behaves like a weak half metal (metallic in one spin channel and almost zero gap in other, see SM\cite{supplement}).

Taking a hint from the spin alignment of different atoms in DO$_3$ disordered structure, we have modelled a FM/AFM interface structure involving Co, Fe and V atoms. This magnetic structure is shown in Fig. S7 of SM,\cite{supplement} which clearly resembles a FM/AFM interface embedded within a host of FM matrix, supporting the spin-valve behavior in CFVS. This magnetic structure is energetically close (within 2 meV) to the lowest energy Type-I FM structure and has a net moment of 2.3 $\mu_B$/f.u.

{\it Summary and Conclusion:}
Here, we report a new quantum material, CoFeVSb, which is predicted to be a potential candidate for spintronics and thermoelectric applications. Using a combined theoretical and experimental study, we predict CoFeVSb to host various interesting multifunctional properties such as room temperature spin-valve nature, large power factor, high spin polarization. The system crystallizes in a cubic structure with small amount of DO$_3$ disorder, as confirmed by both normal/synchrotron XRD and EXAFS measurements. Both ab-initio calculations and resistivity data confirms half metallic  nature, leading to high spin polarization. Observation of double hysteresis loop, asymmetric magnetoresistance and stability of FM/AFM interface from ab-initio calculations strongly suggests the room temperature spin-valve behavior. Hall measurements show anomalous behavior, dominated by the intrinsic Berry phase contribution. Further, large power factor and ultra low lattice thermal conductivity confirm CFVS to be a potential candidate for TE applications as well.
In conclusion, coexistence of so-many interesting properties in a single material opens up new opportunities for numerous applications such as spincalorics, thermoelectrics etc.

{\it Acknowledgments:} JN acknowledges the financial support provided by IIT Bombay in the form of fellowship. Authors thank Dr. Velaga Srihari, ECXRD beamline, BL-11, Indus-2, RRCAT for carrying out anomalous x-ray diffraction measurements. AA acknowledges DST-SERB (Grant No. CRG/2019/002050) for funding to support this research.
\bibliographystyle{apsrev4-1}
\bibliography{references}
\pagebreak
\clearpage

\begin{center}
\textbf{\large Supplementary Material for ``CoFeVSb: A Promising Candidate for Spin Valve and Thermoelectric Applications''}
\end{center}

\renewcommand{\bibnumfmt}[1]{[S#1]}
\renewcommand{\citenumfont}[1]{S#1}


Here, we present the details of experimental synthesis, different measurement tools, and ab-initio computation. We have demonstrated the details of the surface morphology, XRD refinement, EXAFS, synchrotron XRD, EDS data, resistivity and Hall data for CoFeVSb. We present further results on ab-initio calculations.



\maketitle

\section{Experimental details} Polycrystalline samples of CoFeVSb were synthesized using an arc melting system in a high purity Ar environment of the stoichiometric quantities of constituent elements having a purity of at least 99.99\%. To stabilize the desired phase, samples were annealed for 1 week at $900^{\circ}$ C in sealed quartz tubes, followed by quenching in ice-water. For the structural studies, X-ray diffraction (XRD) patterns at RT were taken using Cu-K$\alpha$ radiation with the help of Panalytical X-pert diffractometer. Crystal structure analysis was done using FullProf Suite software.\cite{rodriguez1993recent} EXAFS measurements were carried out at the Energy-Scanning EXAFS beamline (BL-9) in transmission mode at the INDUS-2 Synchrotron Source (2.5 GeV, 200 mA) at the Raja Ramanna Centre for advanced Technology (RRCAT), Indore, India \cite{poswal2014commissioning,basu2014comprehensive} 
The ratio of the measured compositions turns out to be $1.0:0.96:1.1:1.2$, indicating the almost desired stoichiometry. Magnetization measurements at various temperatures were obtained using a vibrating sample magnetometer (VSM) attached to the physical property measurement system (PPMS) (Quantum design) for fields up to 70 kOe. Temperature and field dependent resistivity measurements were carried out using a PPMS (DynaCool) employing the electrical transport option (ETO) in the four-probe method (10 mA current at 18 Hz frequency). Hall measurements were also carried out using the same setup, with the van der Pauw method under the same current and frequency. Thermoelectric power (TEP) was measured using differential dc sandwich method in an in-house setup in the temperature range of 4–300 K.\cite{sharath2008simple}\\
As XRD cannot provide complete structural details of this system because of the fact that the constituents are near neighbors in the periodic table, with almost identical atomic scattering amplitudes,
EXAFS was used. For the EXAFS measurements the Energy-Scanning EXAFS beamline (BL-9) was operated in the energy range of 4-25 KeV. The beamline optics consisted of a Rh/Pt coated 
collimating meridional cylindrical mirror and the collimated beam reflected by the mirror was monochromatized by a Si(111) (2d=6.2709 Å) based double crystal monochromator (DCM). 
The second crystal of the DCM was a sagittal cylinder, which was used for horizontal focusing while a Rh/Pt coated bendable post mirror facing down was used for vertical focusing 
of the beam at the sample position. Three ionization chambers (300 mm length each) were used for data collection in transmission mode, one ionization chamber for measuring incident 
flux (I$_0$), second for measuring transmitted flux (I$_T$) and the third for measuring EXAFS spectrum of a reference metal foil for energy calibration. Appropriate gas pressure and 
gas mixture were chosen to achieve 10-20\% absorption in the first ionization chamber and 70-90\% absorption in the second ionization chamber to improve the signal to noise ratio. 
Rejection of the higher harmonics content in the X-ray beam was performed by second mirror. The absorption coefficient was obtained using the relation:
\begin{equation}
\renewcommand{\theequation}{S\arabic{equation}}
I_T = {I_0} e^{-\mu x}
\label{eq:exaf}
\end{equation}                                                            	
Synchrotron XRD measurement was carried out using the synchrotron wavelength $\lambda = 0.6525 \AA$. Synchrotron based powder x-ray diffraction measurements were performed on well-ground powder samples at Extreme Conditions Angle Dispersive/Energy dispersive x-ray diffraction (EC-AD/ED-XRD) beamline (BL-11) at Indus-2 synchrotron source. Measurements were carried out in capillary mode and capillary was rotated at $\sim$ 150 rpm to reduce the orientation effects. Desired wavelength for ADXRD diffraction experiments was selected from the white light from the bending magnet using a Si(111) channel cut monochrometer. The monochromatic beam was then focused on to the sample with a Kirkpatrick-Baez mirror or K-B mirror. A MAR345 image plate detector (which is an area detector) was used to collect 2-dimensional diffraction data. Sample to detector distance and the wavelength of the beam were calibrated using NIST standards LaB$_6$ and CeO$_2$. Calibration and conversion/integration of 2D diffraction data to 1D, intensity vs 2$\theta$, was carried out using FIT2D software \cite{hammersley1996two}.\\ \\

Surface morphology of the bulk sample was analyzed by atomic force microscopy (AFM) and magnetic force microscopy (MFM) using NanoScope Multimode-IV Veeco, Digital Instruments. Transmission electron microscopy (TEM) images and selected area electron diffraction (SAED) patterns have been recorded using FEG-TEM (Philips CM200) on a powder sample.

  \begin{figure}[t]
 \renewcommand{\thefigure}{S\arabic{figure}}
\centering
\includegraphics[width=8cm,height=5cm]{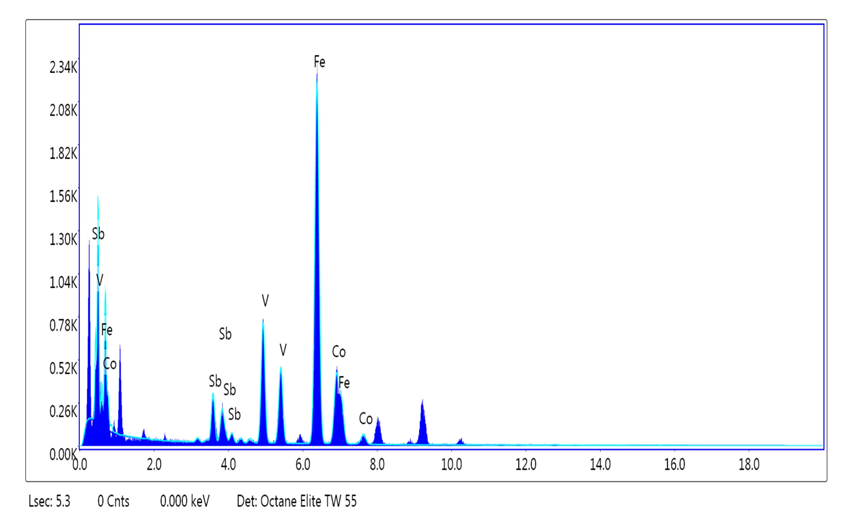}
\caption{For CoFeVSb, EDS spectrum of the constituent elements obtained by EDS.}
\label{fig:eds}
\end{figure}
\section{Computational details}
 Ab initio calculations were performed using spin polarized density functional theory (DFT)\cite{hohenberg1964inhomogeneous} implemented within Vienna ab initio simulation package (VASP) \cite{kresse1996efficient,kresse1996efficiency,kresse1993ab} with a projected augmented-wave (PAW) basis.\cite{kresse1999ultrasoft} The exchange-correlation potential due to Perdew, Burke, and Ernzerhof (PBE) \cite{perdew1996generalized} was used within the generalized gradient approximation (GGA) scheme. Brillouin zone integration was done within the tetrahedron method using a $24\times24\times24$ k-mesh. A plane wave energy cut-off of 500 eV was used for all the calculations. All the structures were fully relaxed with total energies (forces) converged to values less than 10$^{-6}$ eV (0.01 eV/\AA). In order to further evaluate the potential of CFVS for thermoelectric (TE) applications, we simulated the lattice thermal conductivity ($\kappa_L$) using Ab Initio Conductivities (AICON) code\cite{fan2020aicon}. AICON is the modified Debye-Callaway (DC) model, which calculates $\kappa_L$ by including the contribution from both acoustic as well as optical phonon branches scaled by their specific heat ratio.
The mode velocities, Debye temeperature, mode Gruneisen parameters are some of the key quantities used as the input parameters for this model which were calculated from the phonon band structure using Phonopy code\cite{togo2015first}. This requires density functional perturbation theory (DFPT)\cite{baroni2001phonons} calculations, as implemented within the VASP\cite{kresse1996efficiency,kresse1996efficient,kresse1993ab} package. Fig. 4(b) in the manuscript shows the phonon band structure for ordered CoFeVSb. The mode velocities ($\nu$) for acoustic modes are calculated by taking the slope of the band corresponding to vibrational mode at $\Gamma$ point. The Debye temperature ($\theta$) is obtained from the maximum frequency corresponding to the vibrational mode. In addition to $\nu$ and $\theta$, this model for the lattice thermal conductivity also requires the mode Gruneisen parameters ($\gamma_i$).  $\gamma$ is a measure of the degree of anharmonicity of the lattice. Higher the $\gamma$ value, more anharmonic the lattice will be. $\kappa_L$ is defined as:
\begin{equation}
\renewcommand{\theequation}{S\arabic{equation}}
\kappa_L = R_A \frac{\kappa_{LA}+\kappa_{TA}+\kappa_{TA'}}{3} + R_O \kappa_{O}
\end{equation}
where $LA$ denotes longitudinal acoustic, $TA$/$TA'$ denote two transverse acoustic branches and $O$ denotes the optical phonon branches. $R_A = c_V^{aco}/\left(c_V^{aco}+c_V^{opt}\right)$ and
$R_O = c_V^{opt}/\left(c_V^{aco}+c_V^{opt}\right)$
are the specific heat ratio for acoustic and optical branches respectively, and $c_V^i$ are the specific heat of the respective phonon branches ($i$=acoustic/optical).

 \section{Experimental Results}
 \subsection {Composition analysis} 
 The energy-dispersive spectra (EDS) for CoFeVSb sample is shown in the Fig. \ref{fig:eds}. The ratio of the constituent elements was found to be $1.0:0.96:1.1:1.2$, which agrees with the desired composition of the sample.

\subsection {Surface morphology analysis} 
\begin{figure}[t]
 \renewcommand{\thefigure}{S\arabic{figure}}
\centering
\includegraphics[width=8cm,height=4cm]{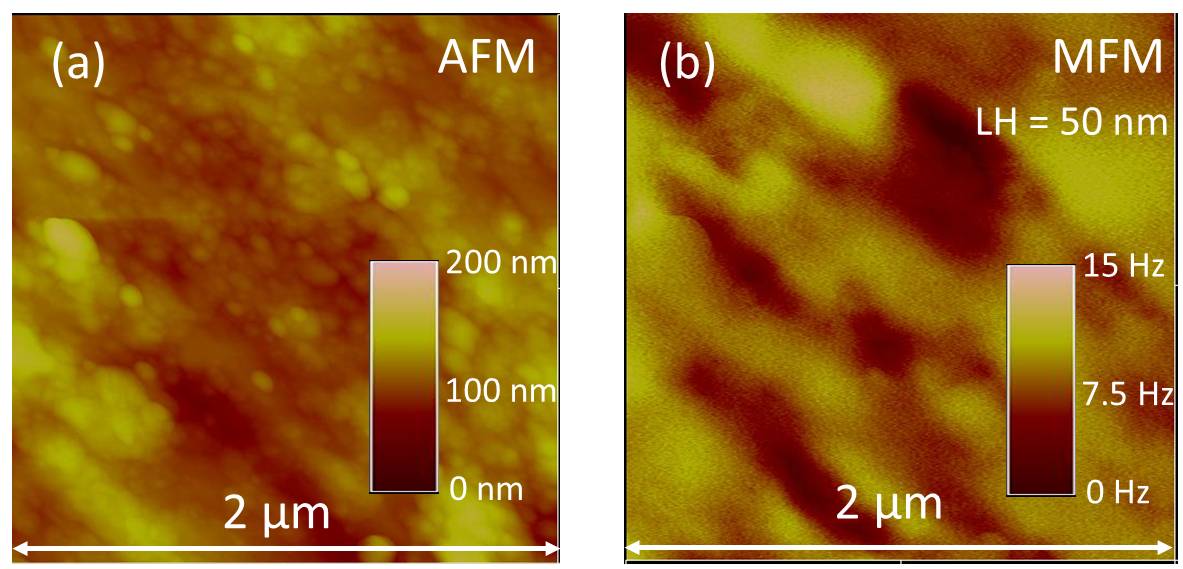}
\caption{For CoFeVSb (a) topographic (AFM) and (b) magnetic (MFM) surface images.}
\label{fig:tem}
\end{figure}

\begin{figure}[t]
\renewcommand{\thefigure}{S\arabic{figure}}
\centering
\includegraphics[width=8.7cm,height=4.0cm]{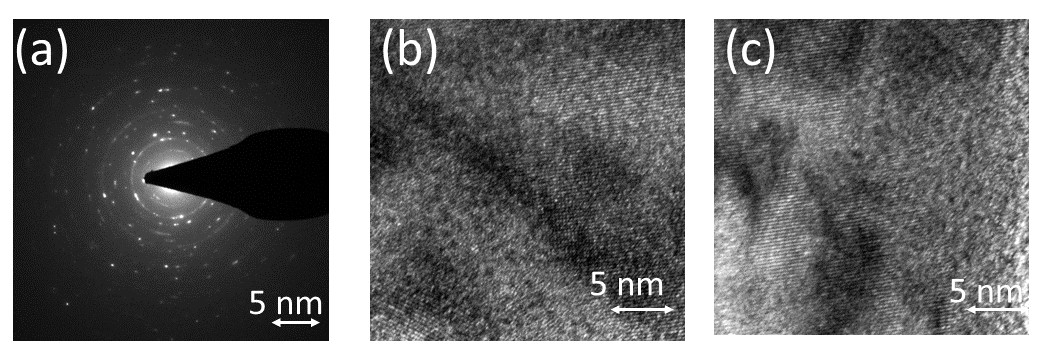}
\caption{For CoFeVSb (a) SAED pattern and (b-c) HR-TEM images.}
\label{fig:mfm}
\end{figure}

Fig. \ref{fig:tem}(a-b) shows the AFM and MFM surface images of the sample. The root mean square (RMS) roughness is found
to be 15 nm when $2\times2$ $\mu m^2$ area of the surface was scanned. \textcolor{black}{It is well known that in MFM the phase and frequency of the oscillating cantilever is mapped while scanning the sample surface. A repulsive magnetic force results in the shifting of the resonance curve towards a higher frequency accompanied by an increase in phase shift, which gives rise to the bright contrasts, while an attractive magnetic force gradient gives rise to dark contrasts in MFM image.} Bright and dark contrast regions in MFM surface image indicate different magnetic domains of competing magnetic phases as shown in Fig. \ref{fig:tem}(b). \textcolor{black}{Hence, our MFM data strongly supports the existence of competing ferromagnetic (FM)/antiferromagnetic (AFM) interface structures within
a host of FM matrix.} Fig. \ref{fig:mfm} (b-c) shows the high resolution transmission electron microscopy (HR-TEM) image for CoFeVSb while its SAED pattern is shown in Fig. \ref{fig:mfm} (a).

\subsection{Crystal Structure}  
\subsubsection { XRD}
 The crystal structure of CoFeVSb can be seen as four interpenetrating fcc sublattices with Wyckoff positions 4$a$, 4$b$, 4$c$ and 4$d$. In general, for a quaternary XX$'$YZ alloy, there exists three energetically non-degenerate structural configurations (keeping Z-atom at 4$a$-site). They are
\begin{itemize}
\item X at 4$c$, X$'$ at 4$d$ and Y at 4$b$ site (Type I)
\item X at 4$b$ , X$'$ at 4d and Y at 4c site (Type II)
\item X at 4$c$ , X$'$ at 4$b$ and Y at 4$d$ site (Type III)
\end{itemize}
For a detailed XRD analysis, the structure factor with the configuration I (Z-atom at 4$a$-site, X at 4$c$, X$'$ at 4$d$, and Y at 4$b$ site) can be expressed as 
\begin{equation}
\renewcommand{\theequation}{S\arabic{equation}}
F_{hkl} = 4(f_Z + f{_Y}e^{{\pi}i(h+k+l)} + f{_X}e^{\frac{{\pi}i}{2}(h+k+l)} + f_{X'}e^{-\frac{{\pi}i}{2}(h+k+l)}).
\label{eq:sfactor}
\end{equation}
where $(hkl)$ are the miller indices. Here $f_X$, $f_{X'}$, $f_Y$, and $f_Z$ are the atomic scattering factors for $X$, $X'$, $Y$, and $Z$ atoms respectively. For super lattice reflections, i.e. for (111) and (200), the structure factor takes the form,
\begin{equation}
\renewcommand{\theequation}{S\arabic{equation}}
F_{111} = 4[( f{_Y} - f_Z ) - i( f{_X} - f_{X'})]
\label{eq:sfactor111}
\end{equation}
\begin{equation}
\renewcommand{\theequation}{S\arabic{equation}}
F_{200} = 4[( f{_Y} + f_Z ) - ( f{_X} - f_{X'})]
\label{eq:sfactor200}
\end{equation}
 Fig. \ref{fig:xrd-CFVS} shows XRD pattern along with the refinement for 
 configuration-I with 12\% disorder between Co-V (X-Y) and 12\% disorder between Fe-V (X$'$ -Y) atoms.
Rietveld refinement was also done by fitting configurations II and III, but those did not fit well. We have also performed refinement considering several  degrees of antisite disorder 
such as 5, 10, 15, 20 \% between Co-V and Fe-V sites for all the three configurations (I, II and III). However, the best fit with the lowest $\chi^2$ (2.01) was 
found with 12.5\% antisite disorder between Co-V sites and 12.5\% antisite disorder between Fe-V sites in the configuration-I. 

\begin{figure}[b]
\renewcommand{\thefigure}{S\arabic{figure}}
\centering
\includegraphics[width=1.0\linewidth]{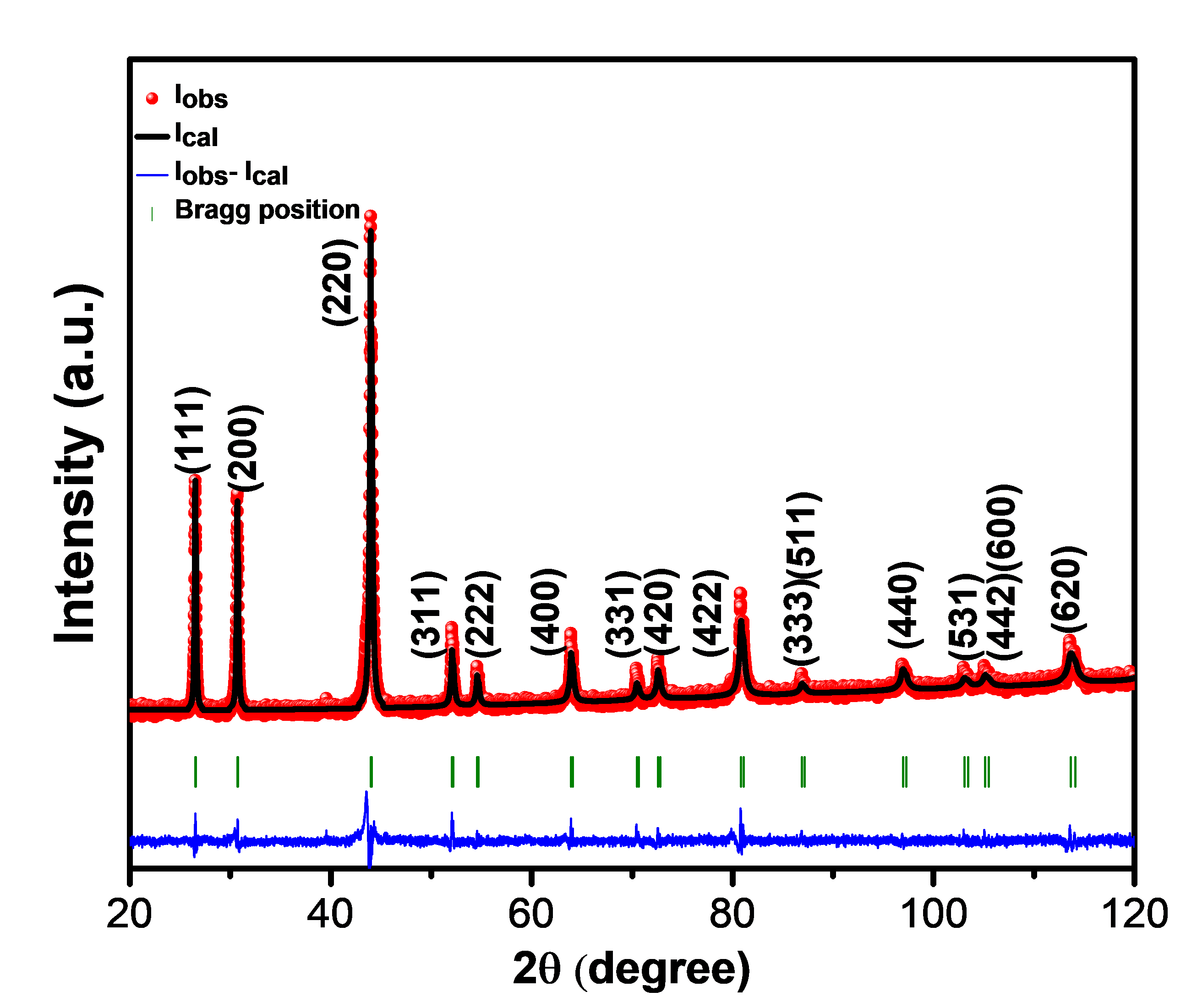}
	\caption{For CoFeVSb (a) Room temperature powder XRD pattern, including the Rietveld refined data for type I configuration with DO$_3$ disorder, i.e. 12.5\% disorder between 
	Co-V and 12.5\% disorder between Fe-V pair of atoms.}
\label{fig:xrd-CFVS}
\end{figure}

 \subsubsection {EXAFS}
 \begin{table*}[t]
 \renewcommand{\thetable}{S\arabic{table}}
\centering
\caption{Bond length, coordination number and disorder factor obtain by EXAFS fitting for CoFeVSb at Co, Fe and V K-edges. $\star$ denotes the contribution from antisite disorder.}
\begin{tabular} {|c|c|c|c||c|c|c|c||c|c|c|c|}
\hline
\multicolumn{4}{|c|} {\textbf{\emph{Co K-edge}}} & \multicolumn{4}{|c|}  {\textbf{\emph{Fe K-edge}}} & \multicolumn{4}{|c|} {\textbf{\emph{V K-edge}}} \\
\hline
\textbf{Path} & \textbf{R ($\AA$)} & \textbf{N} & \bm{$\sigma^2$} & \textbf{Path} & \textbf{R ($\AA$)} & \textbf{N} & \bm{$\sigma^2$} &\textbf{Path} & \textbf{R ($\AA$)} & \textbf{N} & \bm{$\sigma^2$}\\
\hline
Co-V & 2.54$\pm$0.02 & 4 & 0.0076$\pm$0.0016 & Fe-Sb &2.53$\pm$0.02 & 4 & 0.0095$\pm$0.0015 & V-Co & 2.54$\pm$0.02 & 4 & 0.0076$\pm$0.0016\\
\hline
Co-Sb &2.54$\pm$0.02 & 4 & 0.0013$\pm$0.0011 & Fe-V & 2.51$\pm$0.02 & 4 & 0.0164$\pm$0.0013 & V-Fe & 2.51$\pm$0.02 & 4 & 0.0164$\pm$0.0013 \\
\hline
Co-Fe & 2.93$\pm$0.02 & 6 & 0.0094$\pm$0.0033 & Fe-Co & 2.93$\pm$0.02 & 6 & 0.0094$\pm$0.0033& V-Sb &2.94$\pm$0.03 & 6 & 0.0079$\pm$0.0028\\
\hline
$\star$ Co-Co/Fe$_{(V)}$ &2.54$\pm$0.02 & 4 & 0.0075$\pm$0.0015 & $\star$ Fe-Co/Fe$_{(V)}$ & 2.45$\pm$ 0.03 & 4 & 0.0021$\pm$0.0010 &  $\star$ V-V$_{(Co/Fe)}$ & 2.49$\pm$0.03 & 8 & 0.0012$\pm$0.0010
\\
x=0.25 & & &   &  x=0.35   & & &  & x=0.25 & & & \\
\hline
\end{tabular}
\label{tab:exafs}
\end{table*}
 In order to take care of the oscillations, absorption spectra $\mu (E)$ has been converted to absorption function $\chi (E)$ defined as\cite{poswal2014commissioning}
\begin{equation}
\renewcommand{\theequation}{S\arabic{equation}}
\chi(E)=\frac{\mu(E)-\mu_0(E)}{\Delta \mu_0 (E_0)}
\end{equation}
where $E_0$ is the absorption edge energy, $\mu_0 (E_0)$ is the bare atom background and $\Delta \mu_0 (E_0)$ is the step in $\mu (E)$ value at the absorption edge. The absorption coefficient $\chi (E)$ can also be expressed in terms of wave number ($\chi (k)$) using the following relation,
\begin{equation}
\renewcommand{\theequation}{S\arabic{equation}}
k = \sqrt{\frac{2m(E-E_0)}{\hbar^2}}
\end{equation}
here, $m$ is the mass of electron. $\chi (k)$ is weighted by $k^2$ to amplify the oscillation at higher $k$.  The $k^2 \chi (k)$ functions are Fourier transformed to generate the $\chi(R)$ spectra in terms of the real distances from the center of the absorbing atom. The programme available within IFEFFIT software package has been used for EXAFS data analysis.\cite{newville1995analysis} This includes background reduction and Fourier transform to derive the $\chi(R)$  vs. $R$ spectra from the absorption spectra (using ATHENA software). This helps to generate the theoretical EXAFS spectra starting from an assumed crystallographic structure and finally fit the experimental data with the theoretical spectra using ARTEMIS software. Co, Fe and V K-edge are fitted simultaneously with common fitting parameters. This simultaneous fitting reduces the number of independent parameters and enhances the statistical significance of the fitting model. The structural parameters (atomic coordination and lattice parameters) of CoFeVSb used for simulation of the theoretical EXAFS spectra are taken from XRD results.
Table \ref{tab:exafs} shows the detailed EXAFS fitting results. The first peak in Fourier transform EXAFS spectrum at V K-edge (see Fig. 2(d) of the manuscript) arises from the contribution of V-Co, V-Fe and V-Sb coordination shells along with the antisite disorder contribution from V-V$_{Co}$/V$_{Fe}$ (V$_{Co}$ means V at Co site and V$_{Fe}$ means V at Fe site). The first peak at Fe K-edge arises from the contribution of Fe-Sb, Fe-V and Fe-Co. Relatively smaller (2.27 $\AA$) Fe-Fe/Co bond distance is due to the possible presence of metallic Fe, which corroborates with the x-ray absorption near edge structures (XANES) result. The first peak at Co K-edge arises from the contribution of Co-V, Co-Sb and Co-Fe coordination shells. The antisite scattering contribution Co-Co/Fe$_{V}$ is found at 2.54 $\AA$.
  \subsubsection {Synchrotron XRD}
 To support our normal XRD results, we have carried out synchrotron XRD measurements for CoFeVSb. Fig. 1 of the manuscript shows room temperature synchrotron XRD pattern of CoFeVSb, 
 which gives a similar set of reflection peaks as that of the normal XRD. From synchrotron XRD pattern, it appears that the (111) peak is much more intense and prominent as compared to 
 the (200) peak. This becomes more clear from synchrotron XRD data than normal XRD. To understand the DO$_3$ disorder in this system in more detail, we have analysed the relative 
 intensities of the superlattice reflection peaks (111) and (200). The calculated value of relative intensities of the superlattice peaks ((I$_{111}$/I$_{200}$)) was found to be 1.35, which further confirms the presence of DO$_3$ ordering in this alloy.
Similarly, relative values of (I$_{111}$/I$_{220}$) and (I$_{200}$/I$_{220}$) are found to be 0.37 and 0.27 respectively. These values agree fairly well with those obtained from 
the normal room temperature XRD data. Hence, synchrotron data fully supports the results of structural analysis as obtained using the normal XRD data. 

\subsection {Magnetic Properties}
\begin{figure}[b]
\renewcommand{\thefigure}{S\arabic{figure}}
\centering
\includegraphics[width= 8cm,height=6.5cm]{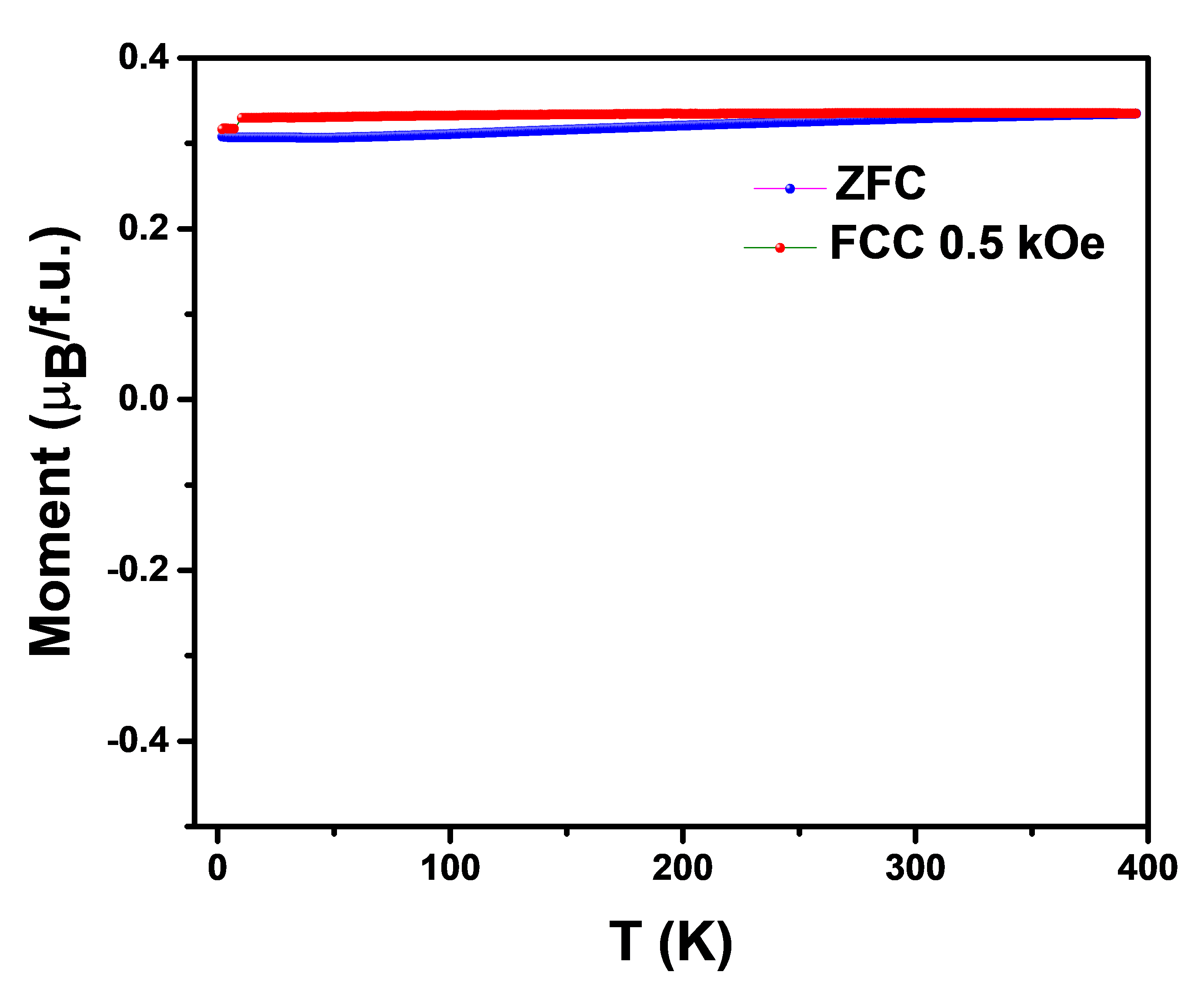}
	\caption{For CoFeVSb, M vs. T curve in zero-field cooled (ZFC) and field cooled (FC) modes in H=0.5 kOe.}
\label{fig:mt-CFVS}
\end{figure}

Fig. \ref{fig:mt-CFVS} shows M vs. T curves for CoFeVSb. It clearly shows high T$_C$ ($>$400 K) for the present compound.
\subsection {Transport Properties}
\begin{table*}[t]
\renewcommand{\thetable}{S\arabic{table}}
\centering
\caption{Power law fitting parameters for $\rho$ vs. T data in different T-range (at zero field) for CoFeVSb.}
\begin{tabular}{l c c c c |}
\hline \hline
 Region & Temperature Range  &  Fitting Equation & Fitting Parameters \\ \hline \hline
I  & 3 $<$T$<$ 13 K &  $ \rho(T)$= $\rho_0$ - B T$^{0.5}$        &  B =- 0.32  $\mu \Omega$-cm K$^{-0.5}$   \\
   & & &   $\rho_0$=96.6  $\mu \Omega$-cm  \\
     \hline
II &   14 $<$T$<$ 50 K &  $ \rho(T)$= $\rho_{0}$ + CT$^{2}$       &   C=$1\times 10^{-4}$  $\mu \Omega$-cm K$^{-2}$ \\
    & & &  $\rho_0$=97.55  $\mu \Omega$-cm  \\ \\
   & &  $ \rho(T)$= $\rho_{0}$ + $\rho_1 T$+ $\rho_2 T^{2}$  & $\rho_1$= 0.02 $\mu \Omega$-cm K$^{-1}$   \\
   & & & $\rho_2$= 0.0001 $\mu \Omega$-cm K$^{-2}$   \\
    && &   $\rho_0$=97.8  $\mu \Omega$-cm  \\ \\
    & &  $ \rho(T)$= $\rho_{0}$ + $D T^{3}$  & D =$6.4\times 10^{-5}$ $\mu \Omega$-cm K$^{-3}$   \\
    && &   $\rho_0$=97.8  $\mu \Omega$-cm  \\
     \hline
III  & 60 $<$T$<$ 310 K &  $ \rho(T)$= $\rho_0^{\prime}$ + AT$^{n}$        &  $n=1.03$   \\
    & &   &A=0.13  $\mu \Omega$-cm K$^{-n}$ \\    
     & &&   $\rho_0^{\prime}$=92.1  $\mu \Omega$-cm  \\
\hline \hline
\end{tabular}
\label{tab:RT}
\end{table*}
 To evaluate the contribution of different scattering mechanisms, we utilize the general power law fitting along with different models in various T-regimes. In the low T regime (region-I),
 T$^{0.5}$ power law fits well. In the region-II, we have tried to fit different models e.g. T$^{2}$, T$^{3}$ and combination of linear and quadratic behavior as shown in the inset (ii) of Fig. 4(a) of the manuscript. The general power law has the following form :
\begin{equation}
\renewcommand{\theequation}{S\arabic{equation}}
\rho (T) = \rho_0 + AT^n
\label{eq:rho-T}
\end{equation}
 Fitting parameters considering various models in different T-regimes are shown in Table \ref{tab:RT}. In the regime-I, there is a clear hump around 15 K and the resistivity data 
 follow $T^{\frac{1}{2}}$ behavior well, indicating the possibility of weak localization at low temperature (T $<$15 K). In the regime-II, the data do not fit well with the T$^{2}$ ($\rho (T) =  \rho_0 + C T^2 $) or T$^{3}$ ($\rho (T) = \rho_0 + D T^3 $) models throughout the T-range. Best fitting for this region was obtained considering $\rho (T) = \rho_0 + \rho_{1} T+ \rho_2 T^2$.

\begin{figure}[b]
\renewcommand{\thefigure}{S\arabic{figure}}
\centering
\includegraphics[width=8.5cm,height=4.5cm]{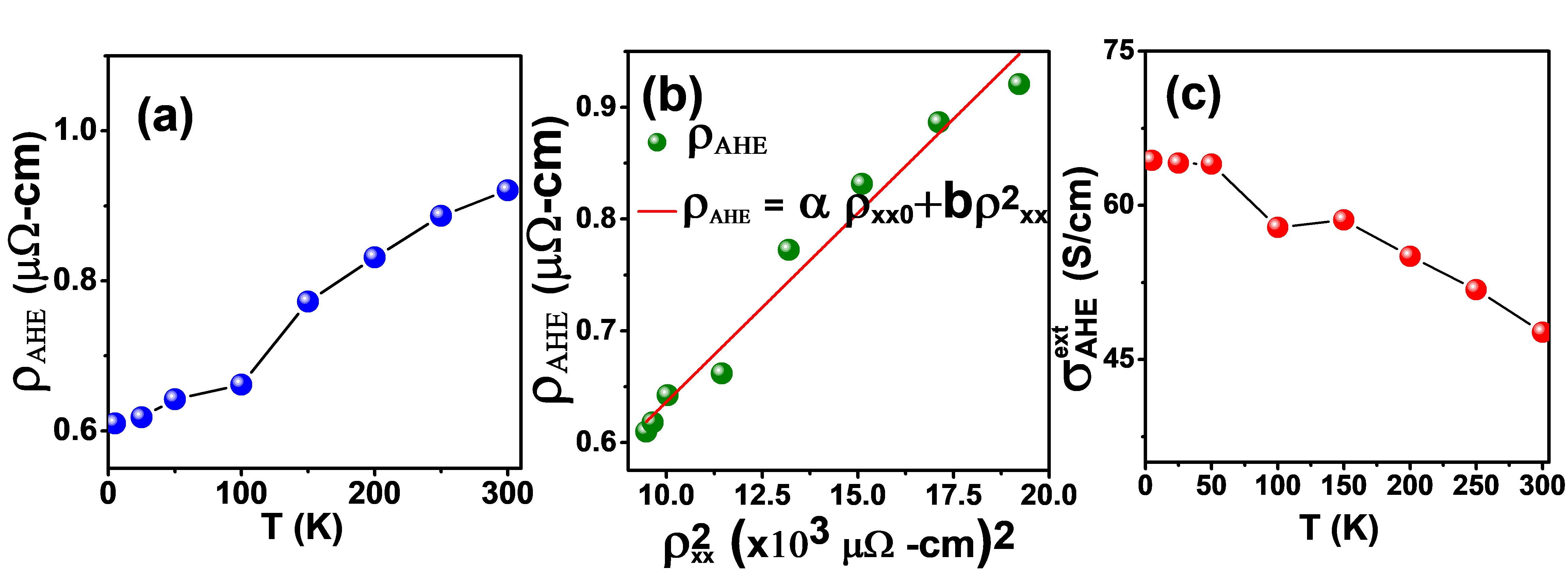}
\caption{For CoFeVSb, (a) and (b) show the variation of $\rho_{AHE}$ with temperature and $\rho_{AHE}$ with $\rho_{xx}^{2}$ along with the fit to Eq. \ref{eq:AHE}. (c) T-dependance of extrinsic contribution of $\sigma_{AHE}$.} 
\label{fig:Hall-CFVS}
\end{figure}

\begin{figure}[b]
\renewcommand{\thefigure}{S\arabic{figure}}
\centering
\includegraphics[width=7cm,height=5cm]{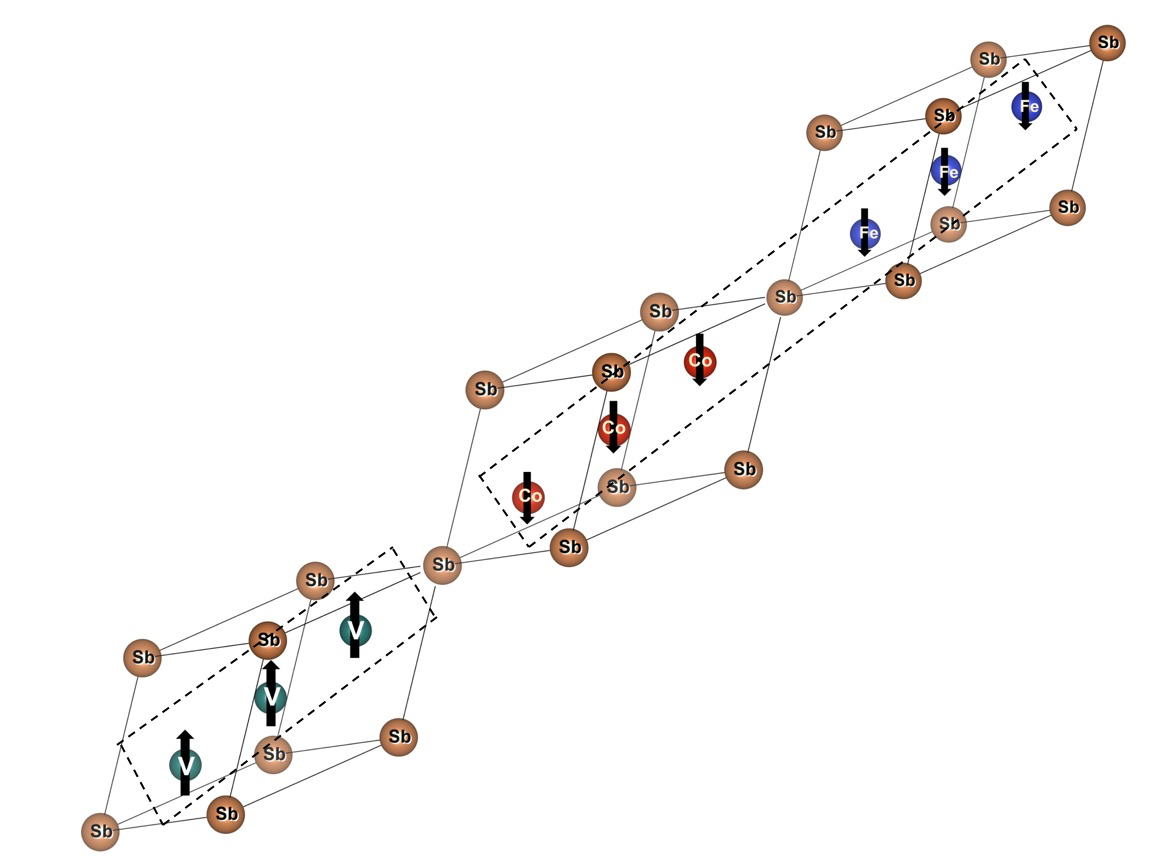} 
\caption{For CoFeVSb, FM/AFM interface structure involving Co, Fe and V atoms embedded within a host of FM matrix.} 
\label{fig:fm-afm}
\end{figure}

\subsection{Hall measurements}
 Figure \ref{fig:Hall-CFVS}(a) shows the temperature dependance of $\rho_{AHE}$ for CoFeVSb, which has a similar behavior as that of longitudinal resistivity($\rho_{xx}$). AHE is generally composed of two components: intrinsic and extrinsic\cite{RevModPhys.82.1539}. The transverse velocity of Bloch electrons in an ideal magnetic crystal gives rise to intrinsic AHE, which depends on the band structure of the material only. Karplus and Luttinger first proposed the mechanism of such intrinsic contribution to AHE, which was reconstructed later in terms of Berry phase.\cite{PhysRev.95.1154,PhysRevB.59.14915} The extrinsic mechanism, on the other hand, arises from the spin-orbit coupling-induced asymmetric scattering of electrons near impurity sites, and this can be classified into (i) Skew scattering \cite{smit1958spontaneous} and (ii) side-jump scattering.\cite{PhysRevB.2.4559}
To find out the intrinsic and extrinsic contributions to AHE, we have used a scaling model reported by Tian et al.,\cite{PhysRevLett.103.087206} according
to which $\rho_{AHE}$ can be expressed as:
\begin{equation}
\renewcommand{\theequation}{S\arabic{equation}}
\rho_{AHE}=a \rho_{xx0} + a^\prime  \rho_{xx0}^{2} + b  \rho_{xx}^{2}
\label{eq:AHall}
\end{equation}
Separating the T-independent and dependent terms, Eq. \ref{eq:AHall} takes the form
\begin{equation}
\renewcommand{\theequation}{S\arabic{equation}}
\rho_{AHE}=\alpha \rho_{xx0} + b  \rho_{xx}^{2}
\label{eq:AHE}
\end{equation}
where $\alpha$ =$ a + a'$ $\rho_{xx0}$ is the T-independent term representing the extrinsic part coming from the skew scattering and side jump impurity scattering, while $b$ is the contribution due to intrinsic part. $\rho_{AHE}$ vs. $\rho_{xx}^{2}$ along with the fit to Eq. \ref{eq:AHE} is shown in Fig. \ref{fig:Hall-CFVS} (b). From the fitting, $\alpha$ and $b$ values turn out to be $0.31 \times 10^{-2}$ and $33.8$  S cm$^{-1}$ respectively.

For a better understanding of extrinsic and intrinsic contributions to AHE, we consider the anomalous Hall conductivity $\sigma_{AHE}$, which can be written as\cite{PhysRevLett.103.087206}:
\begin{equation}
\renewcommand{\theequation}{S\arabic{equation}}
\sigma_{AHE}= \sigma_{int} + \sigma_{ext} = \sigma_{int} + \sigma_{sk} + \sigma_{sj}
\label{eq:sigma-ahe}
\end{equation}
where, $\sigma_{int}$ is the intrinsic Karplus-Luttinger term while $\sigma_{sk}$ and $\sigma_{sj}$ represent the skew scattering and side jump contributions respectively.
Considering Eq. \ref{eq:AHall} and \ref{eq:AHE}, Eq. \ref{eq:sigma-ahe} can be expressed as:
\begin{equation}
\renewcommand{\theequation}{S\arabic{equation}}
\sigma_{AHE}=-( a \sigma_{xx0}^{-1} + a^\prime  \sigma_{xx0}^{-2}) \sigma_{xx}^{2} - b  =  \alpha  \sigma_{xx0}^{-1} \sigma_{xx}^{2} + b
\label{eq:sigma-ahe}
\end{equation}
where $\sigma_{xx0} = \frac{1}{\rho_{xx0}}$ is the residual conductivity. Figure \ref{fig:Hall-CFVS}(c) shows the T dependance of $|{\sigma_{ext}}|$.

\begin{figure}[t]
\renewcommand{\thefigure}{S\arabic{figure}}
\centering
\includegraphics[width=7.5cm,height=5.0cm]{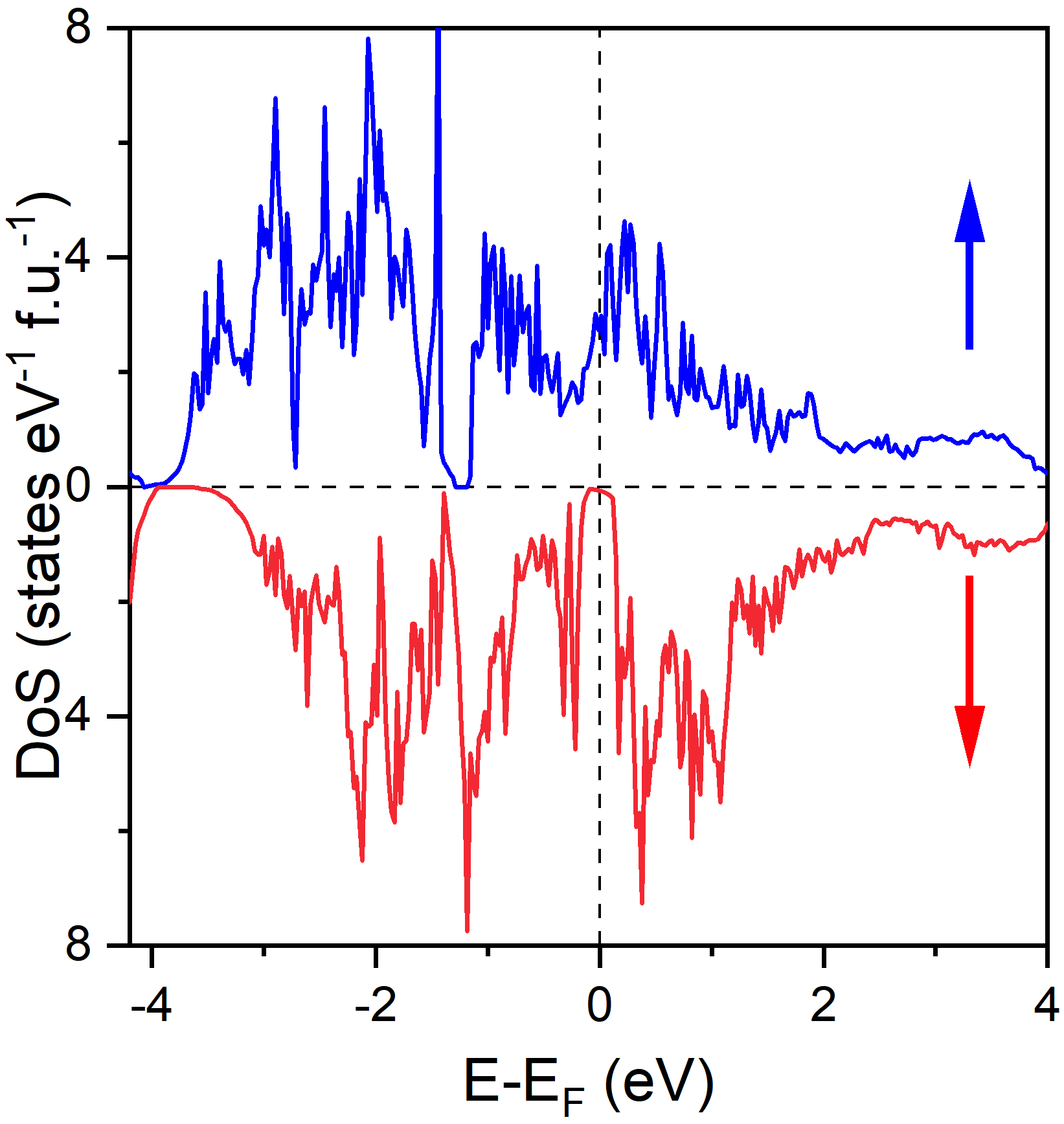}
\caption{Spin polarized density of states of CoFeVSb considering partial DO$_3$ disorder.} 
\label{fig:dos-do3}
\end{figure}
\begin{table}[t]
\renewcommand{\thetable}{S\arabic{table}}
\centering
\caption{ Relaxed lattice parameters ($a_0$), atom projected and total magnetic moments (in $\mu_B$), and relative energies ($\Delta E$) of Type I, Type II and Type III configurations of CoFeVSb, calculated within the GGA approximation.} 
\begin{tabular}{l c c c c c c}
\hline \hline
Type&  $a_0$ (\AA) $ \ $  &  $m^{\mathrm{Co}}$ & $\ $ $m^{\mathrm{Fe}}$ $\ $  &  $m^{\mathrm{V}}$  & $\ $ $m^{\mathrm{Total}}$ $ \ $ & $\Delta E$(eV/f.u.) \\ \hline
I   &  5.9   & 1.04 	& 	 1.13  	& 	 0.84 	& 3.0		&  0   \\
II  &  6.1  & 1.2		& 1.9	& 	-1.3 	&  1.78		& 0.7    \\
III    &  6.0  &  1.056  	&     2.6 	&	 -1.11 	& 2.55 		&  0.36   \\
\hline \hline
\end{tabular}
\label{tab:theory-CFVS}
\end{table}



\end{document}